# Default risk modeling beyond the first-passage approximation: Position-dependent killing.

Yuri A. Katz

Standard & Poor's

55 Water Str., New York, NY 10041

Diffusion in a linear potential in the presence of position-dependent killing is used to mimic a default process. Different assumptions regarding transport coefficients, initial conditions, and elasticity of the killing measure lead to diverse models of bankruptcy. One "stylized fact" is fundamental for our consideration: empirically default is a rather rare event, especially in the investment grade categories of credit ratings. Hence, the action of killing may be considered as a small parameter. In a number of special cases we derive closed-form expressions for the entire term structure of the cumulative probability of default, its hazard rate and intensity. Comparison with historical data on global corporate defaults confirms applicability of the model-independent perturbation method for companies in the investment grade categories of credit ratings and allows for differentiation between "microscopic" models of bankruptcy in the high-yield categories.

1. **Introduction.**

Stochastic modeling of the time evolution of complex systems has a long history in natural and social sciences [1 2]. Remarkably, the first probabilistic formulation of the Brownian motion was related to valuation of stock options by Bachelier [3]. The price of an option is derived from the present price of the underlining equity and depends upon its stochastic evolution over time. Bachelier postulated that at any given time the future of the stock price is entirely determined by its present value. This assumption is the same as the one used by Einstein in his derivation of the diffusion coefficient of Brownian particles in terms of microscopic fluctuations of their positions [4]. Now it is known as the Markov postulate. Both Bachelier and Einstein assumed that the time evolution of the system – the stock price and the location of a Brownian particle, respectively – follows the diffusion process. Further development of the Bachelier's model lead to realization that a better description of equity markets can be achieved under the assumption that the logarithm of the stock price is evolving in accordance with the generalized Wiener process with a non-zero drift (the geometric Brownian motion, see, e.g., [5]). This approximation lays in the foundation of the Black-Scholes-Merton framework, which is commonly used to price equity derivatives [6] and in the "structural" models of default [7,8,9].

Events on stock and credit markets are interrelated, complex, and fundamentally *indeterminate*. Nevertheless, similarly to natural sciences, an adequate representation of phenomena in financial economy may be obtained by using certain phenomenological constructs augmented by measurements of relevant "macroscopic" variables. For instance, in survival analysis and reliability theory the likelihood of destruction is characterized by the relevant time-dependent hazard rate function [10,11]. It determines the risk of failure as a function of time, conditional on not having happened previously. In essence, it describes "how the past affects the future" and, hence, is the main characteristic of a generally non-Markovian evolution of a system. Similarly, in finance the hazard rate function characterizes the risk of default, i.e., failure to pay financial obligations, and is the key attribute of the phenomenological "reduced-form" framework [12 13]. The description of default via the hazard rate leads to closed-form pricing formulae that can be calibrated to fit observable prices of credit risk sensitive securities. If the recovery rate and the risk-free yield curve are known, this procedure enables measurements of the market-implied probability of default (PD), which can be used to estimate prices of illiquid securities bearing similar credit risk [14 15]. Another important advantage of the reduced-form approach is an adequate account for short-term difference in yields of risk-bearing defaultable corporate bonds and risk-free credit instruments of the corresponding maturity. These credit spreads reflect market's uncertainty regarding a company's ability to serve its debt that, contrary to forecasts of traditional structural models, is not zero even at short times to maturity.

As any phenomenological approach, the reduced-form framework is applicable without any explanation of the "microscopic" causes of default. Similarly to laws of thermodynamics, which were put into practice well before the statistical physics, it may be used without any microscopic justification. Yet, even an approximate understanding of a default mechanism and its relationship to empirical regularities are very important. Only a self-consistent theory allowing for *derivation* of the hazard rate function from some microscopic modeling assumptions may validate a priori specifications that have been employed, e.g., in a number of "jump-to-default" equity derivative models [16 17 18]. Verification of these conjectures is impossible within the reduced-form approach. Similarly to traditional structural models, such a theory should be based on clear economic micro-foundations, provide insight and intuition, and allow for differentiation between alternative mechanisms. The new approach, however, should evade the deterministic nature of traditional structural models, which fully pre-specify the future.

It is remarkable in this context that almost a century ago Smoluchowski has derived the closed-form expression for a hazard (recombination) rate function of an irreversible chemical reaction assisted by diffusion of reagents [19]. He established the fundamental connection between the conditional probability density function (pdf) describing a relative position of microscopic particles and the phenomenological hazard rate function. The problem was solved within the first-passage approximation, which implies that a diffusive path of reagents stops at any encounter. This model has been frequently used in natural and social sciences, including the Black-Scholes-Merton option-theoretical framework (see, e.g., [20,21] and references therein). Formally, it requires the solution of the diffusion equation with the absorbing boundary condition, which relates the pdf of a particle at any future point in time exactly to its location at the initial point in time. Therefore, in the first-passage approximation the outcome of a continues-time diffusion process with known initial conditions is *pre-determined*. In 1949, Collins and Kimball have extended this model [22]. They took into account that in reality not every encounter of reagents leads to reaction and introduced the intrinsic rate constant to characterize the efficiency of recombination at the contact. This parameter reflects incomplete knowledge regarding an elementary act of chemical reaction. The extended Smoluchowski's model does not fully prescribe the future of any particular encounter. However, even an extended model has a serious drawback related to the assumed contact character of recombination, which leads to zero reaction (hazard) rate if the point of contact is inaccessible. This inconsistency was addressed in the theory of long-distance luminescence quenching, which incorporates the position-dependent dissipative term into transport equations [23]. Nowadays, different flavors of this theory are used to describe a variety of processes in physics, chemistry, and molecular biology [24 25].

It is the purpose of this paper to apply the basic ideas and formalism developed in these studies to valuation of a default risk. We assume that default is characterized by the position-dependent (*elastic*) killing term introduced into the Fokker-Planck equation (FPE), which in accordance to the structural approach describes the geometric Brownian motion of the pre-default firm's assets value. Introduction of an elastic killing measure naturally brings the concept of incomplete information and indeterminacy of default into the Black-Sholes-Merton framework. Our approach draws on insights gained from the works of Duffie and Lando [26], Jarrow and Protter [27], Giesecke [28] and other authors [29 30 31 32 33 34 35] that are based on the postulate of the fundamental market uncertainty regarding the key financial parameters of a firm. Systematic misreporting of the key financial parameters in the pre-default time has been clearly demonstrated recently by Podobnik et al. [36]. Moreover, these authors have shown that similarly to long-distance energy or electron transfer, default may happen at any given point in "space" (leverage ratio) and time.

One "stylized fact" is fundamental for our quantitative consideration: empirically bankruptcy is a rather rare event, especially in the investment grade categories of credit ratings [37]. Hence, the *action* of killing may be considered as a small parameter of the problem. To apply the perturbation method we transform the FPE with the position-dependent dissipative term into the integral equation. The latter is formally equivalent to the Feynman integral equation for quantum propagators [38] with a scattering potential replaced by an elastic killing term. The new formulation of the problem facilitates the robust perturbation expansion relating the Green functions of the FPEs with and without dissipation. Different assumptions regarding the transport coefficients, initial conditions, and the elasticity of the killing measure lead to diverse models of default, including many of structural and incomplete information models already known in the literature. In a number of special cases we derive closed-form expressions for the entire term-structure of the cumulative PD, its hazard rate and intensity. We establish the limits of validity of models that exploit the assumption of the time-independent hazard rate. Comparison between obtained formulae and historical data on average global corporate defaults [37] clearly demonstrates i) the applicability of the model-independent perturbation method for companies with high credit quality; ii) reveals excellent fitting capabilities of the extended Black-Cox (EBC) model [39] in all categories of credit ratings; iii) allows for differentiation between "microscopic" models of bankruptcy for companies in the speculative grade categories.

**II. The framework.**

To fix ideas and simplify notations, consider the single causal state variable $x = -\ln R$ that represents the company's position in "space". Here $R$ is the firm's leverage defined as the ratio between firm's total liabilities $L$ and its aggregate asset's value $V$. Generalizations including many dimensions will be straightforward once the relevant basic results are well understood. For simplicity let us consider stochastic trajectories only with $R \leq 1$ or equivalently $x \geq 0$. The relevant conditional cumulative survival probability $\Omega(t|x_0)$ is associated with the conditional pdf $p(x,t|x_0,0) \equiv p(x,t|x_0)$ as follows

$$\Omega(t|x_0) = \int_0^\infty p(x,t|x_0)dx. \qquad (2.1)$$

It is well known that $\Omega(t|x_0)$ is connected to the main phenomenological characteristics of default: the cumulative hazard function $H(t,x_0)$ and the hazard rate function $h(t,x_0)$ (see, e.g., Refs.[9 - 13]):

$$H(t,x_0) = -\ln\Omega(t,x_0) = \int_0^t h(\tau, x_0)d\tau. \qquad (2.2)$$

It follows from this expression that the hazard rate function represents the instantaneous cumulative PD, $P(t,x_0) = 1 - \Omega(t,x_0)$, at a given time $t$ conditional on the survival up to $t$:

$$h(t,x_0) = -\Omega^{-1}(t,x_0)\frac{d\Omega(t,x_0)}{dt} = \Omega^{-1}(t,x_0)\frac{dP(t,x_0)}{dt}. \qquad (2.3)$$

Consequently, the intensity $\lambda(t,x_0)$ of a new default occurring is defined as

$$\lambda(t,x_0) = h(t,x_0)\Omega(t,x_0) = \frac{dP(t,x_0)}{dt}. \qquad (2.4)$$

We would like to stress here that the hazard rate function and the intensity of default approximately concur only if $\Omega(t,x_0) \approx 1$, i.e., for very low PDs, $P(t,x_0) \ll 1$. Although obvious, this important distinction between the main characteristics of a default process is often overlooked in the literature. It follows from Eq.(2.3) that the conditional cumulative survival probability satisfies the non-Markov kinetic equation:

$$\frac{d\Omega(t,x_0)}{dt} = -h(t,x_0)\Omega(t,x_0), \qquad \Omega(0,x_0) = 1, \qquad (2.5)$$

which is analogous to the one derived by Smoluchowski in 1917 [19].

Expressions (2.1), (2.2) and equations (2.3) - (2.5) describe relationships between empirically relevant "macroscopic" characteristics of a default process. They constitute the core of the *model-independent* reduced-form approach. On the other hand, Eq.(2.1) implies that the Markov process is used to model the irregular behavior of the state variable *x*. Following the traditional structural modeling approach, let us further assume that the time behavior of *V* is determined by the geometric Brownian motion. Hence, under the physical (objective) measure the stochastic dynamics of *x* satisfy the Ito differential equation with generally position-dependent, time invariant transport coefficients [5, 8, 12, 13]

$$dx = a(x)dt + \sigma(x)dW. \qquad (2.6a)$$

Here $a(x) = \mu(x) - \sigma^2(x)/2$ is the effective drift, $\mu(x)$ is the expected rate of change in a firm's leverage ratio, $\sigma(x)$ is the volatility of a company's assets value, and *W* represents the standard Wiener process. Consequently, the conditional pdf $p(x,t|x_0,0)$, satisfies the following one-dimensional FPE

$$\partial_t p(x,t|x_0) = -\nabla_x J(x,t|x_0), \qquad (2.6b)$$

where $J(x,t|x_0)$ is the probability density flux, which is defined at any point *x* as

$$J(x,t|x_0) = a(x)p(x,t|x_0) - D(x)\nabla_x p(x,t|x_0), \qquad (2.7)$$

with the diffusion coefficient $D(x) := \sigma^2(x)/2$.

Now let us introduce an elastic killing measure, $k(x,t) \geq 0$, which describes the probability per unit time and unit length that the stochastic trajectory is terminated at a given point *x* at a given time *t*. The resulting computational framework can be formulated in terms of the FPE with the position-dependent dissipation term (see also Refs.[24 – 26]):

$$\partial_t p(x,t|x_0) = -\nabla_x J(x,t|x_0) - k(x,t) p(x,t|x_0) \ . \tag{2.8}$$

If the conditional pdf $p(x,t|x_0)$ is known, the key phenomenological characteristics of a default process can be easily calculated, see Eqs.(2.1) – (2.4). In the new formulation of the problem two processes determine bankruptcy: destruction due to the contact with the default barrier, which is determined by the relevant boundary condition and distant position-dependent killing in the bulk. Formally, the 1$D$ FPE with the elastic dissipation term, Eq.(2.8), is analogous to the one considered in the jump-to-default equity pricing models [17, 18]. In these reduced-form models, the dependency of the hazard rate function and the relevant intensity of default upon "space" and time were postulated [17, 18]. In the framework presented here these key characteristics of a bankruptcy process can be derived from some deeper set of assumptions including the capital structure of a company. Thereby, it may capture an interesting, in general, non-linear linkage between events on stock and credit markets, prices of equity options and the likelihood of default.

The FPE (2.8) should be solved with the relevant initial and boundary conditions, which depend on the assumptions of the "microscopic" model of default. For example, in the popular CreditGrades model [30] it is assumed that

$$p(x,0|x_0) = \varphi(x - x_0) \ , \tag{2.9}$$

where $\varphi(x - x_0)$ is some normalized function, $\int \varphi(x - x_0) dx_0 = \int \varphi(x - x_0) dx = 1$, which reflects market's indeterminacy in the initial value of $x$ (see, e.g., Refs.[30, 33, 39]). To simplify notations hereafter we are not showing the limits of integration over *x* from 0 to infinity. Usually it is supposed that

$$J(x,t \mid x_0)_{|x \to +\infty} = 0 \ . \tag{2.10}$$

Starting with Merton's structural model, the common approach in the credit risk literature is based on the assumption that default happens at the *first passage*, i.e., irreversibly and instantly, whenever the diffusive path of *V* hits the *absorbing* default barrier, which is equal to *L* or some lower threshold. This approximation requires zero pdf at the barrier, i.e., $p(0,t \mid x_0) = 0$, whenever $V = L$ or equivalently $R = 1$ and $x = 0$. It follows from Eq.(2.7) that in this case:

$$J(x,t \mid x_0)\big|_{x \to 0_+} = -D(x)\frac{\partial p(x,t \mid x_0)}{\partial x}\bigg|_{x \to 0_+} . \tag{2.11}$$

Default risk models can be straightforwardly extended beyond the first passage approximation if we replace the absorbing boundary condition with the more general *radiation* boundary condition [39]:

$$J(x,t \mid x_0)\big|_{x \to 0_+} = -k_c p(0,t \mid x_0), \tag{2.12}$$

Here $k_c$ represents the local rate constant of default at $x = 0$. Obviously, the probability to survive a contact with the default barrier is negatively correlated with $k_c$. If $k_c = 0$, the boundary is reflecting ("*white*"), $J(x,t \mid x_0)\big|_{x \to 0_+} = 0$, while in the opposite limit, $k_c \to \infty$, it is completely absorbing ("*black*"), see Eq.(2.11).

It immediately follows from Eqs.(2.1), (2.4), (2.8), and (2.10) that generally the intensity of default is determined by the sum of two terms: the probability density flux at the boundary and the aggregate dissipation in the bulk:

$$\lambda(t \mid x_0) = -J(x,t \mid x_0)\big|_{x \to 0_+} + \int k(x,t) p(x,t \mid x_0) dx \qquad (2.13)$$

This expression is exact, very general, and provides the computational bridge between the microscopic model-dependent parameters and phenomenological characteristics of default. For instance, in the first-passage approximation, which corresponds to the absorbing boundary condition at $x = 0$ and $k(x,t) = 0$, Eq.(2.13) in concert with Eq.(2.11) yields

$$\lambda(t \mid x_0) = D(x) \frac{\partial p(x,t \mid x_0)}{\partial x}\bigg|_{x \to 0_+} . \qquad (2.14)$$

This expression was first derived by probabilistic methods in the work of Duffie and Lando [27].

To proceed further it is convenient to reformulate the problem in terms of the first kind Volterra integral equation for the Green function $G(x,y,t)$ of Eq.(2.8) [40]. Applying the Duhamel principal we obtain (see Appendix A)

$$G(x,y,t) = G_0(x,y,t) - \int_0^t d\tau \int dz\, G_0(x,z,t-\tau) k(z,\tau) G(z,y,\tau), \qquad (2.15)$$

$$G(x,y,0) = \delta(x-y). \qquad (2.16)$$

Here $G_0(x,y,t)$ represents the Green function of the FPE (2.6b) without the dissipation term. The general solution of the FPE (2.8) is described by the convolution of $G(x,y,t)$ with the initial condition, Eq.(2.9):

$$p(x,t|x_0) = \int G(x,y,t)p(y,0|x_0)dy = \int G(x,y,t)\varphi(y-x_0)dy. \quad (2.17)$$

Notably, Eq.(2.15) is a close analogue of the Feynman integral equation for quantum propagators [37] with an elastic killing term playing the role of a scattering potential. The chance to be killed in the course of transition from point $y$ to point $x$ is accounted for by the second term on the RHS of Eq.(2.15), which permits the following interpretation. Coordinates $(z,\tau)$ represent the place and time at which the *last attempt* to terminate a stochastic trajectory takes place. Thus, transition from $(z,\tau)$ to the final destination $(x,t)$ cannot be terminated and is described by the $G_0(x,z,t-\tau)$, which is analogous to the propagator of a free particle. Obviously, $G_0(x,y,t)$ depends upon modeling assumptions regarding transport coefficients and boundary conditions. In the next sections we explore different models of default derived from the general framework presented here. We shall see that the formulation of the problem in terms of the integral equation (2.15) provides an efficient computational method, which in some special cases makes possible to derive a number of closed-form expressions for the cumulative PD, the hazard rate function, and the intensity of default.

**III. Risky layer with reflection.**

The computational framework presented in the previous section is free from the deterministic concept of an absorbing default barrier. Intuitively, the strength of an elastic killing measure $k(x,t)$ should increase fast enough in the vicinity of some $x = x_m \approx 0$ ($R_m \approx 1$), so that on the average a stochastic trajectory has a high probability to be terminated within a relatively thin "risky" layer while approaching $x_m$. However, the fortune of a company is not predetermined even within this risky zone. There is merely a much higher likelihood of default when a firm's leverage ratio is close to $R_m$. Indeed, the company may adjust its capital structure and eventually leave the dangerous area. To capture the possibility for a financially distressed company to escape default, we introduce the *reflecting* boundary at $x_m$: $J(x_m, t | x_0) = 0$. Thereby, we restrict the stochastic motion to $x \geq x_m$. It is easy to see that this modeling restriction does not change the structure of general expressions obtained in the previous section. The only required modification is in the lower limit of integration over $x$, which should be equal to $x_m$. Qualitatively, in a course of time the peak of the initial distribution $\varphi(x - x_0)$ at $x = x_0$ moves away from ($a > 0$) or towards to ($a < 0$) the area of maximum destruction at $x = x_m$ and expands due to diffusion as $x(t) \sim \sqrt{D(x)t}$. The diffusion length $\sqrt{D(x)t}$ provides the measure of how far the probability density has propagated in either $x$-direction by diffusion at the instance $t$. Similarly to traditional structural models, the smaller the initial distance to $x_m$ the higher is the likelihood of default. Yet, in our model the fate of a firm is undefined even if $x_0 = x_m$!

Let us multiply Eq.(2.15) by $\varphi(y - x_0)$ and integrate both sides over $x$ and $y$ from $x_m$ to $\infty$. Taking into account that the reflective boundary guarantees the conservation law $\int p_0(x, t | x_0) dx = 1$, it is easy to see that this operation yields

$$\Omega(t\,|\,x_0) = 1 - \int_0^t d\tau \int dx\, k(x,\tau) p(x,\tau|x_0). \qquad (3.1)$$

Formula (3.1) is equivalent to the following definition of the conditional cumulative PD

$$P(t\,|\,x_0) = \int_0^t d\tau \int dx\, k(x,\tau) p(x,\tau|x_0) \qquad (3.2)$$

and leads to

$$\lambda(t\,|\,x_0) = \int k(x,t) p(x,t\,|\,x_0) dx, \qquad (3.3)$$

which is, obviously, consistent with the more general Eq.(2.13). Expression (3.3) immediately yields the following microscopic definition of the hazard rate function:

$$h(t\,|\,x_0) = \frac{\lambda(t,x_0)}{\Omega(t\,|\,x_0)} = \frac{\int dx\, k(x,t) p(x,t\,|\,x_0)}{1 - \int_0^t d\tau \int dx\, k(x,\tau) p(x,\tau|x_0)}. \qquad (3.4)$$

Although rather general, expressions (3.1) – (3.4) reflect certain modeling assumptions regarding the microscopic mechanism of default. In particular, firm's ability to overcome the financial distress is modeled by the reflective boundary condition at $x = x_m$. To complete a model one should additionally define the position-dependency of transport coefficients, the distribution of initial values of a firm's leverage ratio, and the dependency of an elastic killing measure upon $x$ and $t$. In the next sections we implement two alternative models originating from different specifications of the position-dependent dissipation term.

**IV. The "Dirac killing" and the extended Black-Cox model of default.**

Let us assume that the killing measure does not explicitly depend on time and its spatial dependency is determined by the density function *f(x)*:

$$k(x,t) = k\, f(x). \qquad (4.1)$$

Now consider the contact (barrier-like) structural model of default, which assumes that bankruptcy may be triggered if and only if the firm's leverage ratio hits certain predefined value, which can be always set as the origin of the positive axis. In the new formulation of the problem this condition corresponds to the risky layer of zero width, which can be represented by the "Dirac killing" term

$$k(x,t) = k_c\, \delta(x) \qquad . \qquad (4.2)$$

It should be qualitatively clear that, if the reflective boundary lays at $x = 0$ and transport coefficients are constant, the solution of the integral equation (2.15) with the initial condition (2.16) and the kernel (4.2) should be equivalent to the outcome of the EBC model of default, which assumes no dissipation in the bulk and the radiation boundary condition at $x = 0$.

Let us make a quick sanity check here. It is readily seen that substitution of Eq.(4.2) into Eq.(2.15) yields

$$G(x,y,t) = G_0(x,y,t) - k_c \int_0^t d\tau\, G_0(x,0,t-\tau) G(0,y,\tau). \qquad (4.3)$$

The structure of the second term on the RHS of this equation reflects the key assumption of all contact models: a stochastic trajectory can be terminated by multiple attempts, but only at one point – at the barrier. We would like to stress here that the reflective boundary condition $J(x,t \mid x_0)\big|_{x \to 0_+} = 0$ and, hence, a chance to escape default, is implicitly present here via the Green function of the *non-dissipative* FPE $G_0(x,y,t)$. Taking the Laplace transform of both sides of Eq.(4.3) we obtain the algebraic equation

$$\tilde{G}(x,y,s) = \tilde{G}_0(x,y,s) - k_c \tilde{G}_0(x,0,s) \tilde{G}(0,y,s), \qquad (4.4)$$

which after straightforward manipulations yields the closed-form expression for the Laplace transform $\tilde{G}(x,y,s) = \int_0^\infty \exp(-st) G(x,y,t) dt$ of the fundamental solution of the problem in terms of $\tilde{G}_0(x,y,s)$

$$\tilde{G}(x,y,s) = \tilde{G}_0(x,y,s) - k_c \frac{\tilde{G}_0(x,0,s) \tilde{G}_0(0,y,s)}{1 + k_c \tilde{G}_0(0,0,s)}, \qquad (4.5)$$

Thus, if the Laplace transform of the Green's function $G_0(x,y,t)$ of the FPE (2.6b) with the reflecting boundary condition is known, it can be utilized to obtain the solution for a much more complex task. In Appendix B we demonstrate how this procedure works and reproduce the outcome of the EBC model of default for the "sharp" deterministic initial condition, $\varphi(x - x_0) = \delta(x - x_0)$.

Naturally, it takes some time for continues transport to reach the default point. Therefore, for short time horizons, $\sqrt{Dt} \ll x_0$, the Dirac killing model leads to severe underestimation of a risk of default. This conclusion does not depend on the strength of the killing measure or choice of the boundary condition. Forecasts of contact models can be improved, however, if one takes into account the indeterminacy in the distance to the default barrier [30, 33, 39]. Following the CreditGrades model [30], suppose that the initial value of *x* is normally distributed with the mean $<x> = x_0$ and the variance $<(x-x_0)^2> = \delta^2$:

$$\varphi(x-x_0) = \frac{1}{\delta\sqrt{2\pi}} \exp\left[-\frac{(x-x_0)^2}{2\delta^2}\right] \quad . \tag{4.6}$$

In this case, the EBC model of default yields

$$P(t\mid x_0) = \frac{P(t+\tau_\delta \mid x_0 - a\tau_\delta) - P(\tau_\delta \mid x_0 - a\tau_\delta)}{1 - P(\tau_\delta \mid x_0 - a\tau_\delta)}, \tag{4.7}$$

$$h(t|x_0) = \frac{k_c}{\Omega(t+\tau_\delta \mid x_0 - a\tau_\delta)} \left\{ \frac{1}{\sqrt{\pi D(t+\tau_\delta)}} \exp\left[-\frac{(x_0+at)^2}{4D(t+\tau_\delta)}\right] - \frac{2k_c + a}{D} \exp\{[x_0 + at + k_c(t+\tau_\delta)]\frac{k_c}{D}\} \Phi\left[-\frac{x_0 + at + 2k_c(t+\tau_\delta)}{\sqrt{2D(t+\tau_\delta)}}\right] \right\}, \tag{4.8}$$

$$P(t+\tau_\delta | x_0 - a\tau_\delta) = \Phi\left[-\frac{x_0 + at}{\sqrt{2D(t+\tau_\delta)}}\right]$$
$$+ \frac{k_c}{k_c + a} \exp[-\frac{a}{D}(x_0 - a\tau_\delta)] \Phi\left[-\frac{x_0 - a(t + 2\tau_\delta)}{\sqrt{2D(t+\tau_\delta)}}\right] \tag{4.9}$$
$$- \frac{2k_c + a}{k_c + a} \exp\{[x_0 + at + k_c(t+\tau_\delta)]\frac{k_c}{D}\} \Phi\left[-\frac{x_0 + at + 2k_c(t+\tau_\delta)}{\sqrt{2D(t+\tau_\delta)}}\right]$$

Here $\Phi(z) = [1 + erf(z/\sqrt{2})]/2$ denotes the cumulative normal distribution, $erf(z)$ is the error function. It is easy to see that the time-shift, $\tau_\delta = \delta^2/2D$, related to uncertainty in the initial distance to the default barrier leads to the non-zero hazard rate function Eq.(4.8) even at very short times to maturity ($t << \tau_\delta$). Note that expressions (4.7) – (4.9) extend formulas of the CreditGrades model [30] beyond the first-passage approximation. The straightforward reduction to the CreditGrades formula for the cumulative PD presented in Ref.[30] can be achieved if $a = -D$ ($\mu = 0$) and $k_c \to \infty$ in Eq.(4.9). However, notice minor errors in the CreditGrades formulas Eqs.(2.11) and (2.12) of the Ref.[30] (see Ref.[39] for details). Furthermore, the CreditGrades model erroneously forecast a non-zero PD at $t = 0$. Instead, one should use expression (4.7), which properly takes into account the normalization term $P(\tau_\delta | x_0 - a\tau_\delta)$. Despite the apparent smallness of this term we couldn't ignore it at short times. Finally, we would like to stress that the numerator of Eq.(4.8), which represents the intensity of default according to the EBC model, generalizes the renowned result of Rubinstein and Reiner for the density of the first hitting time [41].

**V. The "Gaussian killing" and the perturbation method.**

The general framework developed in sections II and III allows us to go beyond the contact approximation. Consider, for example, the static limit, $D = a = 0$. In this extreme case, traditional structural models forecasts zero PDs. However, it is readily seen from Eqs.(2.8), (2.9), and (3.3) that even in this seemingly simple situation the intensity of default is, in general, non-zero and is determined by the non-trivial convolution

$$\lambda(t|x_0) = \int \varphi(x - x_0) \, k(x) \exp[-k(x)t] dx. \qquad (5.1)$$

Thus, the interplay between different sources of market indeterminacy may result in a complex non-Markovian term-structure of the cumulative PD. Remarkably, even in the case of sharp deterministic initial condition, $\varphi(x - x_0) = \delta(x - x_0)$, Eq.(5.1) yields the non-zero time-dependent intensity of default, $\lambda(t|x_0) = k(x_0) \exp[-k(x_0)t]$. Obviously, the situation becomes much more complex if we take into account that a firm's assets volatility is never zero and the pdf $p(x,t | x_0)$ is always affected by the stochastic transport. Hence, one should find the Green function of the FPE Eq.(2.8) with the position-dependent killing term, which is a very challenging computational task.

One of the key advantages of the formulation of the problem in terms of the integral equation (2.15) is its convenience in finding approximate solutions when the *action* of the killing measure is small:

$$k(x)t << 1 \tag{5.2}$$

The iterative substitution of the left hand side of the integral equation (2.15) back into its right hand side leads to the series, which is analogous to the perturbation expansion in the classical Feynman-Kac formula, with relevant path integrals evaluated as ordinary integrals [37]. In the first non-zero order in this parameter the fundamental solution of the FPE (2.8) can be expressed via the relevant Green function of the non-dissipative FPE $G_0(x,y,t)$ as follows

$$G(x,y,t) = G_0(x,y,t) - \int dz k(z) \int_0^t d\tau G_0(x,z,t-\tau) G_0(z,y,\tau) . \tag{5.3}$$

Note that the second term on the RHS of this expression takes into account that the stochastic trajectory may be killed *anywhere*, but only at *one* attempt [cf. Eqs (2.15) and (4.4)]. It follows from Eqs.(2.17), (3.3), (4.1), and (5.3) that in the lowest order in $k_c t$ the intensity of default takes the form

$$\lambda(t \mid x_0) \approx k_c \int f(x) p_0(x,t \mid x_0) dx . \tag{5.4}$$

This expression reflects the simple qualitative consideration: if the action of destruction is weak, it should not radically disturb the pdf, $p(x,t \mid x_0) \approx p_0(x,t \mid x_0)$, where $p_0(x,t \mid x_0)$ is the pdf of the non-dissipative FPE. Formulae for the second-order approximation that take into account that the stochastic trajectory may be terminated anywhere in $1D$ space at two attempts are presented in Appendix C.

In the contact approximation Eq.(5.4) is reduced to $\lambda(t \mid x_0) \approx k_c p_0(0, t \mid x_0)$, which in the case of constant transport coefficients and the reflecting boundary condition at $x = x_m = 0_+$ immediately gives the following result [see Appendix B, Eq.(B1)]:

$$\lambda(t|x_0) \approx k_c \left\{ \frac{1}{\sqrt{\pi D(t+\tau_\delta)}} \exp\left[-\frac{(x_0+at)^2}{4D(t+\tau_\delta)}\right] - \frac{a}{D} \Phi\left[-\frac{x_0+at}{\sqrt{2D(t+\tau_\delta)}}\right] \right\}, \quad (5.5)$$

where we took into account the uncertainty in the initial condition, Eq.(4.6). Qualitatively if the role of diffusion is small, $Dt << (x_0+at)^2$, the influence of the reflective boundary on the pdf should be also very small. Indeed, in this regime the second term on the RHS of Eq.(5.5), which originates from reflection is negligible in comparison with the first one that represents the unbounded diffusion. Therefore, if $Dt << (x_0+at)^2$, we may further reduce Eq.(5.5) to

$$\lambda(t|x_0) \approx \frac{k_c}{\sqrt{\pi D(t+\tau_\delta)}} \exp\left[-\frac{(x_0+at)^2}{4D(t+\tau_\delta)}\right] . \quad (5.6)$$

Now consider the model with the Gaussian spatial distribution of the killing measure

$$k(x,t) = k(x) = \frac{k_c}{\Delta\sqrt{2\pi}} \exp\left[-\frac{(x-x_m)^2}{2\Delta^2}\right], \quad (5.7)$$

where $\Delta$ represents the width of the risky layer. Similarly to the contact model of default, we assume the total reflection at $x_m$. Contrary to the contact approximation, this model allows to distinguish between two qualitatively different regimes: one where diffusion dominates $Dt >> \Delta^2$ (regime A) and the opposite quasi-static regime $Dt << \Delta^2$ (regime B). Intuitively, in the regime A the elasticity of the killing measure is insignificant. In fact, if we completely ignore the width of the risky layer, $\Delta \to 0$, the "Gaussian killing" model described by Eq.(5.7) reduces to the "Dirac killing", Eq.(4.2). On the other hand, in the regime B, we may expect that elasticity of the killing measure is one of the key factors that determine the conditional pdf $p(x,t|x_0)$. However, as we shall see below, this conclusion is not true if the action of killing is small.

It is easy to see from Eqs.(4.6), (5.4), (5.7), and (B.1) that if the reflecting boundary is at $x = x_m = 0_+$ and $Dt << (x_0 + at)^2$, the account of the non-contact "Gaussian killing" as well as the indeterminacy in the initial values of a firms leverage ratio lead to the following result:

$$\lambda(t|x_0) \approx \frac{k_c}{\sqrt{\pi D(t+\tau)}} \exp\left[-\frac{(x_0+at)^2}{4D(t+\tau)}\right], \qquad (5.8)$$

where $\tau = \tau_\delta + \tau_\Delta$, $\tau_\delta = \delta^2/2D$ and $\tau_\Delta = \Delta^2/2D$. Comparison of this formula with Eq.(5.6) clearly demonstrates that as long as the action of destruction is small and $Dt << (x_0 + at)^2$, the relevant asymptotes of the results obtained in the contact (EBC) and non-contact (Gaussian) models of default coincides (up to the re-definition of the time-shift). We shall return to this important point in the next section. Note that as follows from Eq.(5.8) diffusion is insignificant only for short-term maturities, $t << \tau$, where $\lambda(t|x_0)$ is almost entirely determined by the overlap between two normal distributions described by Eqs.(4.6) and (5.7), $\sim \exp[-x_0^2/(\delta^2 + \Delta^2)]$. Evidently, only in this limit the intensity of default can be time invariant, provided $k_c t << 1$ and $|a|t << x_0$.

In general, the intensity of default and the term structure of the cumulative PD is influenced by the stochastic transport and, hence, is described by the complex non-Markovian kinetics. For relatively large initial distances $|a|t \ll x_0$, the integral of the Eq.(5.8) over time can be calculated in the closed-form, which yields the term structure of the cumulative PD:

$$P(t|x_0) = \frac{k_c}{\sqrt{\pi D}}[F(t+\tau) - F(\tau)] \quad , \qquad (5.9)$$

$$F(u) = \sqrt{u}\exp\left[-\frac{x_0^2}{4Du}\right] + x_0\sqrt{\frac{\pi}{D}}\,\Phi\left[\frac{x_0}{\sqrt{2Du}}\right] \quad . \qquad (5.10)$$

We would like to stress here that since the observable cumulative probability of corporate defaults is rather small, especially in the investment grade categories of credit ratings, the inequality (5.2) might be satisfied for an extended period of time. We shall discuss the actual range of validity of the perturbation expansion in the next section and compare the forecasts of different models of default considered in this paper with historical observations.

**VI. Comparison with empirical data.**

Let us map the results derived in the previous sections to observations. According to recent study conducted by Standard and Poor's [38], the frequency of corporate defaults as a percentage of the total count of global obligors originally rated by S&P was 0.36% in 2007 and rose globally through the crises to 3.99% in 2009. Apparently, there is a significant variation of these numbers across regions, industries, and categories of credit ratings. For example, at the end of 2009 the frequency of defaults was only 0.32% in the investment-grade category (AAA – BBB) versus 9.23% in the speculative-grade category (BB - CCC/C). Note that these numbers were at their highest levels since 2002. Therefore, under the physical measure the inequality (5.2) and the perturbation expansion that leads to the appealingly simple formulae for the intensity of default, Eq.(5.8), and the term-structure of the cumulative PD, Eqs.(5.9) and (5.10), may be valid for relatively long time horizons especially for firms in the investment-grade category.

Let us compare the term structure of the cumulative PDs predicted by these formulae with the one forecasted by the EBC model and the empirical data on global average corporate cumulative defaults [38]. Note that in spite of the long-term observations, 1981 – 2009, some historical data on cumulative defaults have low statistical significance, especially in the investment grade category. For example, throughout the 29-years of observations, only 7 companies that were originally rated 'AAA' and 25 that were rated 'AA' have ever defaulted [38]. Therefore, we exclude from our consideration defaulters that were originally rated in the 'AAA' and 'AA' categories, which accounts for only 1.66% of the total number of corporate defaults.

We recall that $D = \sigma^2/2$ and rewrite Eqs.(5.9) and (5.10) in the form that is more convenient for comparison with empirical data:

$$P(t|x_0) = \frac{\tilde{k}_c}{\sqrt{\pi/2}} \left[ F(t+\tau) - F(\tau) \right] \tag{6.1}$$

$$F(u) = \sqrt{u} \exp\left[ -\frac{(\tilde{x}_0 - \tilde{x}_m)^2}{2u} \right] + \sqrt{2\pi}(\tilde{x}_0 - \tilde{x}_m) \Phi\left[ \frac{\tilde{x}_0 - \tilde{x}_m}{\sqrt{u}} \right] \tag{6.2}$$

where $\tilde{k}_c = k_c/\sigma$, $\tilde{x}_{0,m} = x_{0,m}/\sigma$, $\tau = \tilde{\delta}^2 + \tilde{\Delta}^2$, $\tilde{\delta} = \delta/\sigma$, and $\tilde{\Delta} = \Delta/\sigma$. Now we perform the standard stochastic optimization of the parameters [42] for the root mean square deviation:

$$\rho(\vec{Z}) = N^{-1/2}\sqrt{\sum_{i=1}^{N}\left[P(t_i,\vec{Z}) - P_i^h\right]} \quad . \tag{6.3}$$

Here $P_i^h$ is the actual (historical) cumulative PD reported at $t = t_i$ and $\vec{Z}$ is the vector of parameters. We have used a random search algorithm with the uniform distribution in the interval $[z_n q, z_n / q]$, where $z_n$ is the n-th parameter from the previous set of parameters that lowered $\rho$. The initial set of parameters was obtained from the preliminary fit in Excel. The number of trials $N$ was chosen $10^4$ and we set q = 0.9. In Fig.1 we present the results of the fitting: the term structure of the cumulative PD predicted by the EBC model [43], Eq.(B.7), the outcome of the model with Gaussian killing, Eqs.(6.1) and (6.2), and the empirical data on global average corporate cumulative defaults [38] in different categories of credit ratings initially assigned to defaulted companies (see Ref.[38] for details). The optimized fitting parameters are collected in Table I.

Both the EBC and the Gaussian killing models demonstrate an excellent agreement with the empirical data for defaulters, which initially have the investment-grade credit ratings (A or BBB), see Fig.1 (a). Moreover, there are almost no differences neither between the key parameters ($\tilde{x}_0$ and $\tilde{k}_c$) nor in the values of physical PDs generated by these two very different models of default. Note that in the categories A and BBB of initial corporate credit ratings both models lead to small $k_c = (0.03 \div 0.06)\sigma$ and relatively large $x_0 = (2.35 \div 3.38)\sigma$ to fit the historical observations. Intuitively, these estimates of the strength of the killing measure are in order with our expectations – the better credit quality should be associated with the smaller value of $k_c$ and the larger initial distance to default. It is well known that generally the historical assets volatility $\sigma$ is smaller than the equity volatility $\sigma_E$. For firms in the investment-grade categories $\sigma_E$ is usually less than 30% per annum and, hence, $k_c < 1\%/year$ (see Table I). Therefore, we may conclude that inequality (5.2) should be well satisfied for a long-term horizon (up to 30 years), which may explain the convergence of the EBC and Gaussian killing models of default and the outstanding fit of the simple formulae (6.1) and (6.2) to long-term observations in the investment-grade categories of credit ratings.

Situation is very different for companies that were originally assigned the speculative credit ratings ('BB', 'B', and 'CCC/C'). It is easy to see from Fig.1 (b) and especially Fig.1 (c) that the Gaussian killing model does not allow for a good quantitative agreement with observations in these categories of initial credit ratings. Notably, the deviation between the forecast of this model and observations is growing with the decline of the initial credit rating. The EBC model, on the other hand, perfectly fits the empirical data for all categories of initial credit ratings with $x_0$ growing fourteen times from $0.24\sigma$ for 'CCC/C' to $3.34\sigma$ for 'A' and $k_c$ reducing at the same ratio from $0.42\sigma$ for 'CCC/C' to $0.03\sigma$ for 'A'. Therefore, we may conclude that for companies with low credit quality i) the action of destruction is rather high and we cannot restrict our consideration to the lowest order of perturbation theory in $k_c t$; ii) the width of the risky layer is negligibly small in comparison with the diffusion length, $Dt >> \Delta^2$, (asset's volatility is high) and the contact approximation is valid.

**VII. Summary.**

Our results are rather general and can be applied to valuation of different types of risk in situations where a system simultaneously undergoes stochastic transport and position-dependent destruction. Potential applications range from survival analysis in biological organisms to cellular neurobiology, from an assessment of a likelihood of failures in mechanical systems to valuation of social risks. In this paper we have focused on a risk of a corporate default. This scenario is a very infrequent event. Nevertheless, the loss suffered by creditors in the event of default can be enormous. For instance, the debt amount affected by the bankruptcy of Lehman Brothers, financial institution with 158-year history, was $144.4 billion. Ability to assess the probability of such a catastrophic event is of central importance in the credit risk management (including portfolio risk management) and in the pricing of financial instruments.

Data collected in Ref.[36] clearly demonstrate that companies may exist with high leverage ratios, e.g., $R > 2$ as well as file for bankruptcy protection having more assets than debt, $R < 1$. What is more, it has been shown that defaulters systematically misreport the value of key financial parameters in a pre-default time. Therefore, the true values of $R$ are not available for an external market observer before the bankruptcy filing. By design, the computational framework presented here reflects these findings. Our approach provides the bridge between economic micro-foundations of traditional structural and reduced-form models. Introduction of the finite killing measure enables the powerful model-independent perturbation method, which is impossible in the first-passage approximation. In a number of special cases we derive closed-form expressions for the entire term structure of the cumulative probability of default, its hazard rate and intensity. These formulae enable fast interactive "what-if" analysis.

Based on comparison of the derived term structure of the physical cumulative PD with empirical data on average global corporate defaults, we may conclude that i) the contact EBC model allows for an excellent fitting in both investment and speculative grade categories of initial credit ratings; ii) the weak-killing approximation is well satisfied for long-term horizons (up to 30 years) for companies in the investment-grade categories of credit ratings (above BBB-). For these companies one may expect that under the risk-neutral measure the observed term-structure of CDS spreads should be proportional to the intensity of default described by Eq. (5.8). Thus, careful analysis of empirical data may help to differentiate between the weak-killing and the first-passage approximations, which leads to the result of Rubinstein and Reiner, Eq.(4.8). This study is left for future research.

**Acknowledgement**.

I am grateful to N. V. Shokhirev, V. E. Gluzberg, and A. I. Burshtein for many enlightening discussions and critical comments on a range of kinetic problems dealt with in this paper.

**Table 1**. The optimized fitting parameters for Gaussian and EBC models of default ($x_m = 0_+$).

|  | Gaussian 'CCC/C' | EBC 'CCC/C' | Gaussian 'B' | EBC 'B' | Gaussian 'BB' | EBC 'BB' | Gaussian 'BBB' | EBC 'BBB' | Gaussian 'A' | EBC 'A' |
|---|---|---|---|---|---|---|---|---|---|---|
| $\tilde{x}_0$ | 0.01 | 0.24 | 0.05 | 1.15 | 0.97 | 1.85 | 2.35 | 2.46 | 3.38 | 3.34 |
| $\tilde{k}_c$ | 0.20 | 0.42 | 0.12 | 0.29 | 0.10 | 0.18 | 0.06 | 0.05 | 0.04 | 0.03 |
| $\tilde{a}$ | n/a | 0.34 | n/a | 0.35 | n/a | 0.32 | n/a | 0.38 | n/a | 0.21 |
| $\tau$ | 0.00 | n/a | 0.01 | n/a | 0.09 | n/a | 0.63 | n/a | 0.77 | n/a |

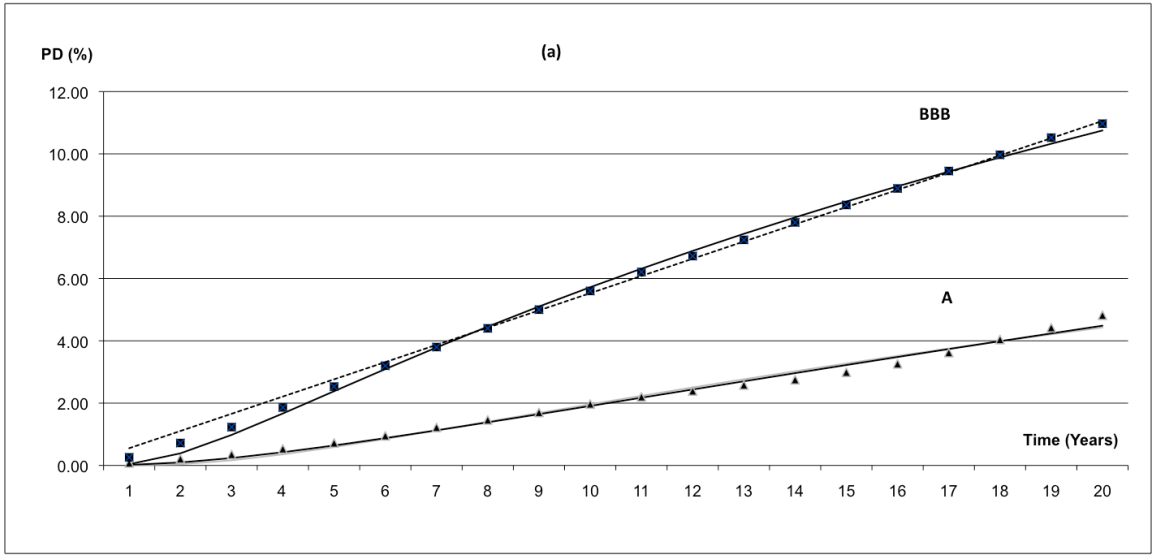

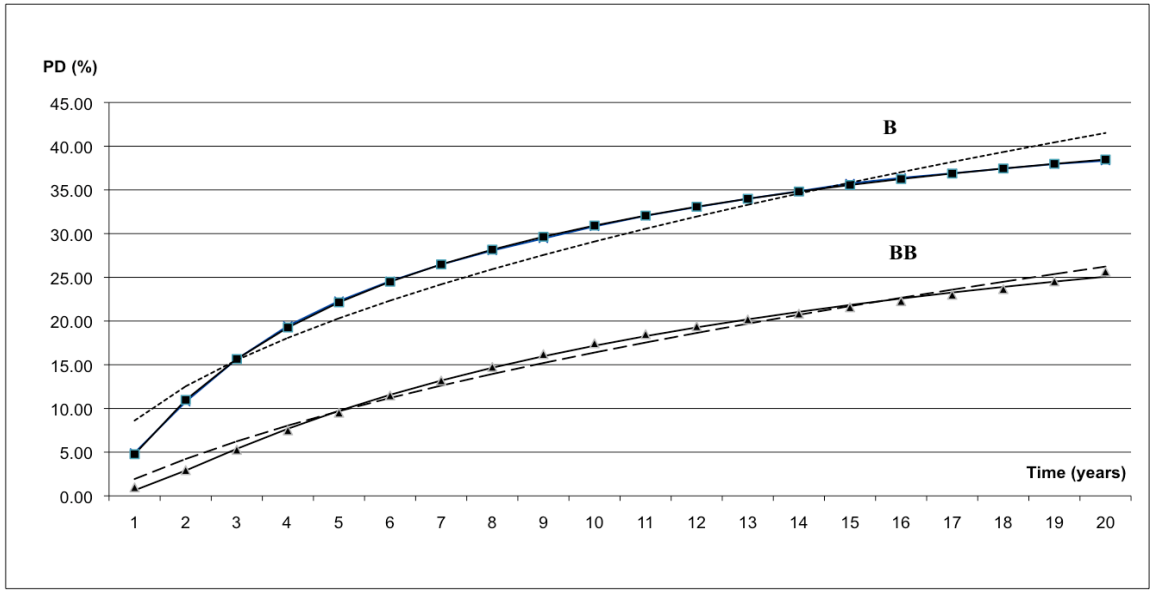

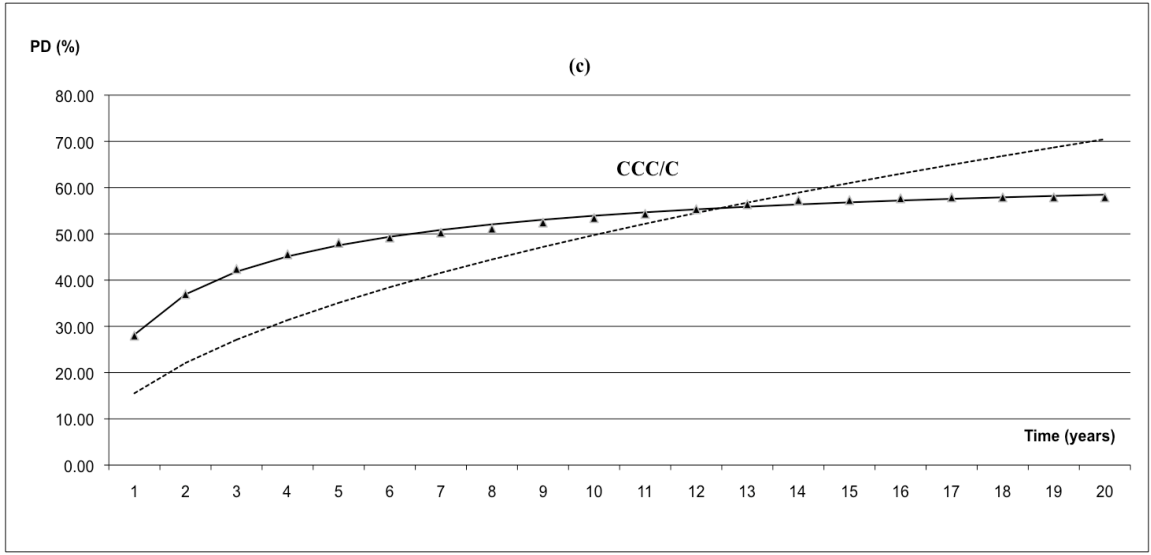

Fig. 1. Observed [37] and calculated here average global corporate cumulative PDs (1981 – 2009). Triangles and squares – empirical data-points; solid lines – outcome of the EBC model; dashed lines – results of the non-contact model of default. (a) Initial credit ratings 'A' and 'BBB'. (b) Initial credit ratings 'BB' and 'B'. (c) Initial credit ratings 'CCC/C'. Note that in the category A of initial ratings the numerical values of the cumulative PDs predicted by both models of default are so close that we cannot distinguish them on the Fig.1 (a).

**Appendix A.**

Apply the linear operator of motion

$$\hat{L}_x = -a(x)\frac{\partial}{\partial x} + D(x)\frac{\partial^2}{\partial x^2} \qquad (A.1)$$

to equation (2.15)

$$\hat{L}_x G(x,y,t) = \hat{L}_x G_0(x,y,t) - \int dz \int_0^t d\tau\, k(z,\tau) G(z,y,\tau) \hat{L}_x G_0(x,z,t-\tau) \qquad (A.2)$$

and take the time derivative of the same

$$\partial_t G(x,y,t) = \partial_t G_0(x,y,t) - \int dz\, G_0(x,z,0) k(z,t) G(z,y,t)$$
$$- \int dz \int_0^t d\tau\, k(z,\tau) G(z,y,\tau) \partial_t [G_0(x,z,t-\tau)] \qquad (A.3)$$

Subtracting (A.2) from (A.3) we have

$$[\partial_t - \hat{L}_x] G(x,y,t)) = [\partial_t - \hat{L}_x] G_0(x,y,t) - \int dz\, G_0(x,z,0) k(z,t) G(z,y,t)$$
$$- \int dz \int_0^t d\tau\, k(z,\tau) G(z,y,\tau)[\partial_t - \hat{L}_x] G_0(x,z,t-\tau)] \qquad (A.4)$$

Now taking into account that by definition $G_0(x,y,0) = \delta(x-y)$ and $[\partial_t - \hat{L}_x]G_0(x,y,t) = 0$ we obtain

$$[\partial_t - \hat{L}_x] G(x,y,t) = -k(x) G(x,y,\tau), \qquad (A.5)$$

which proves that the integral equation (2.15) is equivalent to the extended FPE Eq.(2.8).

**Appendix B.**

The explicit analytical solution of the FPE (2.8) without position dependent sinking term and reflecting boundary condition at $x = 0_+$ was obtained by Smoluchowski for a constant drift $a$ and diffusion $D$ coefficients [44]:

$$G_0(x,y,t) = \frac{1}{2\sqrt{\pi Dt}} \exp\left[\frac{a(x-y)}{2D} - \frac{a^2 t}{4D}\right] \left\{ \exp\left[-\frac{(x-y)^2}{4Dt}\right] + \exp\left[-\frac{(x+y)^2}{4Dt}\right] \right\}$$
$$- \frac{a}{2D} \exp\left[\frac{ax}{D}\right] erfc\left[\frac{at+x+y}{2\sqrt{Dt}}\right] \qquad (B.1)$$

However, the Laplace transform of this expression is not known. Therefore, it is convenient to step back and make the following auxiliary substitution in Eq.(4.3)

$$G(x,y,t) = G^u(x,y,t)\exp\left[\frac{a(x-y)}{2D} - \frac{a^2 t}{4D}\right],$$

$$G_0(x,y,t) = G_0^u(x,y,t)\exp\left[\frac{a(x-y)}{2D} - \frac{a^2 t}{4D}\right], \quad (B.2)$$

which leads to the following expression [cf. Eq.(4.5)

$$\widetilde{G}^u(x,y,s) = \widetilde{G}_0^u(x,y,s) - k_c \frac{\widetilde{G}_0^u(x,0,s)\widetilde{G}_0^u(0,y,s)}{1 + k_c \widetilde{G}_0^u(0,0,s)}. \quad (B.3)$$

The Laplace transform of $G_0^u(x,y,t)$ is easy to obtain

$$\widetilde{G}_0^u(x,y,s) = \frac{1}{2\sqrt{sD}}\exp\left[-\frac{|x-y|}{\sqrt{D}}\sqrt{s}\right] + \frac{1}{2\sqrt{sD}}\exp\left[-\frac{(x+y)}{\sqrt{D}}\sqrt{s}\right]$$

$$- \frac{a}{2D\sqrt{s}\left(\sqrt{s} + \frac{a}{2\sqrt{D}}\right)}\exp\left[-\frac{(x+y)}{\sqrt{D}}\sqrt{s}\right] \quad (B.4)$$

Substitution of Eq.(B.4) into Eq.(B.3) yields

$$\widetilde{G}^u(x,y,s) = \widetilde{G}_0^u(x,y,s) - \frac{1}{\sqrt{D}}\exp\left[-\frac{(x+y)}{\sqrt{D}}\sqrt{s}\right]\left\{\frac{1}{\sqrt{s} + \frac{a}{2\sqrt{D}}} - \frac{1}{\sqrt{s} + \frac{a}{2\sqrt{D}} + \frac{k_c}{\sqrt{D}}}\right\}. \quad (B.5)$$

Substitution of the inverse Laplace transform of Eq.(B.5) into Eq.(B.2) gives the explicit fundamental solution of the problem in the contact approximation, which was first obtained by a different method in Ref.[45]:

$$G(x,y,t) = \frac{1}{2\sqrt{\pi Dt}} \left\{ \exp\left[-\frac{(x-y-at)^2}{4Dt}\right] + \exp\left[-\frac{(x+y-at)^2}{4Dt} - \frac{ay}{D}\right] \right\}$$
$$-\frac{(a+2k_c)}{D} \exp\left[\frac{(a+k_c)(x+k_ct) + k_c y}{D}\right] \Phi\left[-\frac{(a+2k_c)t + x + y}{\sqrt{2Dt}}\right] . \quad \text{(B.6)}$$

Now it is straightforward to obtain the general solution $p(x,t \mid x_0)$ and calculate the cumulative PD and intensity of default. For the "sharp" deterministic initial condition, $\varphi(y - x_0) = \delta(y - x_0)$ we reproduce the main results of the EBC model of default [39]:

$$P(t|x_0) = \Phi\left[-\frac{x_0 + at}{\sqrt{2Dt}}\right] + \frac{k_c}{k_c + a} \exp(-ax_0/D) \Phi\left[-\frac{x_0 - at}{\sqrt{2Dt}}\right]$$
$$- \frac{2k_c + a}{k_c + a} \exp\{[x_0 + (k_c + a)t]\frac{k_c}{D}\} \Phi\left[-\frac{x_0 + (a + 2k_c)t}{\sqrt{2Dt}}\right] , \quad \text{(B.7)}$$

$$\lambda(t|x_0) = k_c \left\{ \frac{1}{\sqrt{\pi Dt}} \exp[-\frac{(x_0 + at)^2}{4Dt}] - \right.$$
$$\left. - \frac{2k_c + a}{D} \exp\{[x_0 + (k_c + a)t]\frac{k_c}{D}\} \Phi\left[-\frac{x_0 + (a + 2k_c)t}{\sqrt{2Dt}}\right] \right\} . \quad \text{(B.8)}$$

**Appendix C.**

The integral equation (2.15) is very convenient for finding the approximate solution in the limit of weak killing, Eq.(5.2). The zero-order approximation is obtained by omitting the second term on the right hand side of Eq. (2.15):

$$G(x,y,t) \approx G_0(x,y,t) . \quad \text{(C.1)}$$

The first-order approximation is obtained by substituting Eq.(C.1) into the right hand side of Eq. (2.15):

$$G(x,x_0,t) \approx G_0(x,y,t) - \int dz \int_0^t d\tau\, G_0(x,z,t-\tau) k(z,\tau) G_0(z,y,\tau). \qquad (C.2)$$

Repeating this process we get the second and higher orders of the perturbation theory. In the second-order approximation

$$\begin{aligned}G(x,y,t) &\approx G_0(x,y,t) - \int dz \int_0^t d\tau\, G_0(x,z,t-\tau) k(z,\tau) G_0(z,y,\tau) \\ &+ \int dz \int_0^t d\tau\, G_0(x,z,t-\tau) k(z,\tau) \int dz_1 \int_0^\tau dv\, G_0(z_1,z,\tau-v) k(z_1,v) G_0(z_1,y,v)\end{aligned} \qquad (C.3)$$

[1] J.-P. Bouchaud, Nature, **455**, 1181 (2008); J.-P. Bouchaud, http://arxiv.org/abs/0904.0805 (2009).

[2] V. M. Yakovenko and J. B. Rosser, Jr., Rev. Mod. Phys. **81**, 1703–1725 (2009).

[3] L. Bachelier, Annales Scientifiques de l'École Normale Supérieure, **3**, 21 (1900).

[4] A. Einstein, Ann. Phys. (Leipzig), **17**, 549 (1905).

[5] J. C. Hull, *Options, Futures, and other Derivatives*. (Pearson, New Jersey, 7th Ed., 2009).

[6] F. Black and M. Scholes, Journal of Political Economy, **81**, 81 (1973); R. C. Merton, Bell Journal of Economics and Management Science, **4**, 141 (1973).

[7] R. C. Merton, Journal of Finance, **29**, 449 (1974).

[8] F. Black and J. C. Cox, Journal of Finance, **31**, 351 (1976).

[9] T. Bielecki and M. Rutkowski, *Credit Risk: Modeling, Valuation and Hedging*. (Springer, Berlin, 2004).

[10] R. E. Barlow and F. Proschan, *A Mathematical Theory of Reliability*, SIAM Classics in Applied Mathematics, v.17 (1996).

[11] O. Aalen, O. Borgan, H. Gjessing, *Survival and event history analysis: a process point of view.* (Springer, 2008)

[12] R. A. Jarrow and S. M. Turnbull, Journal of Finance, **50**, 53 (1995); P. Artzner and F. Delbaen, Mathematical Finance, **5**, 187 (1995); D. Duffie and K. J. Singleton, Review of Financial Studies **12**, 687 (1999); D. Lando, Review of Derivatives Research **2**, 99 (1998)

[13] D. Duffie and K. J. Singleton, *Credit Risk: Pricing, Measurement, and Management* (Princeton University Press, Princeton, 2003).

[14] D. Lando, *Credit Risk Modeling* (Princeton University Press, Princeton, 2004).

[15] For a single entity the actual recovery rate is apparently not known in the pre-default time.

[16] R. C. Merton, Journal of Financial Economics, **3**, 125 (1976).

[17] V. Linetsky, Mathematical Finance, **16**, 255 (2006); P. Carr and V. Linetsky, Finance and Stochastics, **10,** 303 (2006);

[18] V. Linetsky, "Spectral Methods in Derivatives Pricing," in *Handbook of Financial Engineering, Handbooks in Operations Research and Management Sciences*, Elsevier, Amsterdam (2007).

[19] M. v. Smoluchowski, Z. Physikalische Chem., **92**, 129 (1917).

[20] J. Masoliver and J. Perello, Phys. Rev. E, **80**, 016108 (2009).

[21] S. Redner, *A Guide to First-Passage Processes* (Cambridge University Press, Cambridge, 2001).

[22] F. C. Collins and G. E. Kimball, J. Colloid Science, **4**, 425 (1949).

[23] N. N. Tunitskii and Kh. S. Bagdasar'an, Optika i Spektroskopiya, **15**, 303 (1963); S. F. Kilin, M. S. Mikhelashvili, and I. M. Rozman, Optika i Spektroskopiya, **16**, 576 (1964); I. I.Vasil'ev, B. P. Kirsanov, and V. A. Krongaus, Kinetika i Kataliz, **5**, 792 (1964);.

[24] G. Wilemski and M. Fixman, J. Chem. Phys., 58, 4009 (1973); K. M. Hong and J. Noolandi, J. Chem. Phys., **68**, 5163 (1978); A. I. Burshtein, Adv. Chem. Phys., **129**, 105 (2004); D. Holcman, A. Marchewka, and Z. Schuss, Phys. Rev. E, 72, 031910 (2005); A. I. Burshtein, Adv. Phys. Chem., 214219 (2009).

[25] Z. Schuss, *Theory and Applications of Stochastic Differential Equations*, Wiley Series in Probability and Statistics (Wiley, New York, 1980).

[26] D. Duffie and D. Lando, Econometrica **69,** 633 (2001).

[27] R. Jarrow and P. Protter, Journal of Investment Management, **2,** 1 (2004).

[28] K. Giesecke, Journal of Economic Dynamics and Control, **30**, 2281 (2006).

[29] C. Finger, *CreditGrades Technical Document*, RiskMetrics Group (2002).

[30] P. Collin-Dufresne, R. Goldstein, and J. Helwege, Working paper, Carnegie Mellon University, http://www.defaultrisk.com/pp_corr_54.htm (2003).

[31] U. Cetin, R. Jarrow, P. Protter, and Y. Yildirim, Annals of Applied Probability, **14**, 1167 (2004).

[32] K. Giesecke and L. Goldberg, Journal of Derivatives, **12**, 1 (2004).

[33] C. Capuano, IMF Working Paper, No. 08/194 (2008).

[34] W. T. Shaw, Working paper, http://www.defaultrisk.com/pp_other181.htm (2009).

[35] C. Capuano, J. Chan-Lau, G. Gasha, C. Medeiros, A. Santos, and M. Souto, IMF Working Paper, No. 09/162 (2009).

[36] B. Podobnik, D. Horvatic, A. M. Petersen, B. Urosevic, and H. E. Stanely, PNAS, 107, 18325 (2010).

[37] D. Vazza, D Aurora, and N. Kraemer, *Default, Transition, and Recovery: 2009 Annual Global Corporate Default Study And Rating Transitions* (Standard & Poor's, New York, 2010).

[38] Feynman, R. P., and Hibbs, A. R., *Quantum Mechanics and Path Integrals*, New York: McGraw-Hill (1965)

[39] Yu. A. Katz and N. V. Shokhirev, Phys. Rev. E, **82**, 016116 (2010).

[40] A. I. Burshtein, A. A. Zharikov, N. V. Shokhirev, O. B. Spirina, E. B. Krissinel, J. Chem. Phys., **95**, 8013 (1991); A. I. Burshtein, A. A. Zharikov, N. V. Shokhirev, J. Chem. Phys., **96**, 1951 (1992).

[41] M. Rubinstein and E. Reiner, Risk Magazine, April, p.28 (1991).

[42] A. H. G. Rinnooy Kan and G. T. Timmer, Stochastic methods for global optimization, *Am J Math Manag Sci*, **4**, 7 (1984).

[43] To minimize the number of fitting parameters, we assume that $\delta = 0$.

[44] M. v. Smoluchowski, Phys. Zeits., **17**, 557 (1916).

[45] N. Agmon., J. Chem. Phys., **81**, 2811 (1984).